\documentclass[preprint,3p,authoryear]{elsarticle}

\usepackage{graphics}
\usepackage{graphicx}
\usepackage{epsfig}

\usepackage{amssymb}
\journal{New Astronomy}

\begin{document}

\begin{frontmatter}



\title{A photometric study of the neglected eclipsing binaries: \\
V405~Cep, V948~Her, KR~Mon and UZ~Sge}


\author[]{A. Liakos}%\corauthref{cor}}
\ead{alliakos@phys.uoa.gr}
\author[]{P. Niarchos}

\address{Dept. of Astrophysics, Astronomy and Mechanics, University of Athens, GR 157 84 Zografos, Athens, Greece (Hellas)}

\begin{abstract}
CCD multi-band light curves of the neglected eclipsing binaries V405~Cep, V948~Her, KR~Mon and UZ~Sge were obtained and analysed using the Wilson-Deninney code. New geometric and absolute parameters were derived and used to determine their current evolutionary state. V405~Cep, V948~Her and KR~Mon are detached systems with their components in almost the same evolutionary stage. UZ~Sge is a classical Algol system with a tertiary companion around it. Moreover, since the systems V405~Cep, V948~Her and UZ~Sge contain an early type component, their light curves were examined for possible pulsation behaviour.
\end{abstract}


\end{frontmatter}


\section{Introduction}
\label{1}

%general idea
Eclipsing binary systems offer unique conditions to measure fundamental parameters of stars, such as stellar masses, radii and luminosities, which are of great importance in studies of stellar structure and evolution.

The main purposes of this work are: a) the derivation of the geometric and photometric parameters of each system through light curve (hereafter LC) analysis, b) the approximate calculation of the absolute parameters of their components and c) a description of their current evolutionary status. In addition, frequency analysis search was performed for possible pulsational behaviour in the candidate systems for including a $\delta$~Sct component.

%Aim
Concluding, in the present study we selected the following three, previously neglected, systems which are lacking accurate photometric measurements and/or modern LC analysis in order to derive for the first time their main physical characteristics.

%Systems' history
\textbf{V405~Cep:} This system ($B$=8.95~mag, $P$=1.37374$^{\rm d}$) was discovered as a variable by $Hipparcos$ mission \citep{ES97} and has an Algol-type LC. The only available information for this binary concerns its A2 spectral type \citep[cf.][]{WR03,KH01}, while \citet{SO06} included it in their list of candidate systems for including pulsating components.

\textbf{V948~Her:} The system's Algol-type variability was discovered by $Hipparcos$ mission \citep{ES97}. It has a period of 1.27521$^{\rm d}$, a $B$-magnitude of 9.26 and its spectral type is listed as F2 in many catalogues \citep[cf.][]{WR03,BU04}. The system is also candidate for containing a $\delta$~Sct component \citep{SO06}.

\textbf{KR~Mon:} This binary was discovered as a variable by \citet{ZI52}. Its period is 1.15097$^{\rm d}$ and its brightness has a magnitude of 11.7 in $B$-filter. The \textit{GCVS} catalogue \citep{SA11} and the \textit{MK Catalogue of Stellar Spectral Classifications} \citep{SK10} list it as a G3V type star. LCs of the system in $V$ and $I$ filters were obtained by the $ASAS$ project \citep{PO05}, but no analysis has been made so far.

\textbf{UZ~Sge:} The binary nature of this Algol-type system ($B$=11.4~mag, $P$=2.21574$^{\rm d}$) was reported by \citet{GS39}. Its spectral type is given either as A3V \citep{HA84,SK90} or A0V \citep{BD80,BU84}. \citet{BD80} and \citet{SK90} calculated the absolute parameters of its components but they resulted in different values. There are many minimum timings available in the literature since the discovery of the system. \citet{LN08} and \citet{ZA08} analysed its orbital period changes and concluded that, a third body with minimal mass of $\sim$0.7~M$_\odot$ may exist in a wide orbit. \citet{BU09} obtained and analysed $V$ and $R$ LCs of the system and resulted in two possible geometrical configurations with different mass ratios.

\section{Observations and data reduction}
\label{2}

The observations were carried out at the Gerostathopoulion Observatory of the University of Athens from November 2008 to July of 2010. A 40~cm Cassegrain telescope equipped with various CCD cameras and the $BVRI$ photometric filter set (Bessell specification) were used for data collection. In particular, all systems were observed with the ST-10XME CCD, except KR~Mon for which the ST-8XMEi CCD was used.

%Reduction
Standard data reduction processes (i.e. dark and bias removal, flat-field correction) and aperture photometry were applied to the data and differential magnitudes using the software \emph{MuniWin} \citep{HR98} were obtained. The details of observations are given in Table \ref{tab1}, where it is listed: The \emph{nights spent} and the \emph{time span} of the observations, the comparison $C$ and check $K$ stars used with their respective magnitudes in specific bands, and the mean photometric error of the variable-comparison magnitude values (i.e. standard deviation $s.d.$) for each filter used.

%%%%%%%%%%%%%%%%%%%%%%%%%%%%%%%%%%%%%%%%%%%%%%%%%%%%%%%%%%%%%%%%%%%%%%%%%%%%%%%%%%%%%%%%%%%%%%%%%%%%%%%%%%%%%%%%%%%%%%%%%%%%%%%%  Table 1 - Observations log
\begin{table}[h]
\centering
\caption{The photometric observation log.}
\label{tab1}
%\vspace{1mm}%\scalebox{0.8}{
\scalebox{0.97}{
\begin{tabular}{l cc lc cccc }
\hline
System       &  Nights  &  Time span    &    Comparison stars  &      mag      &\multicolumn{4}{c}{s.d. [mmag]}  \\
\cline{4-9}
             & spent    & (Month/Year)  &                      &               & $B$ & $V$ & $R$ & $I$           \\
\hline
V405 Cep     &  5       &    10/09      & $C$: TYC 4516-0540-1 & 9.65$^a$ ($V$)& 2.4 & 2.4 & 2.3 & 5.2           \\
             &          &               & $K$: TYC 4516-1224-1 & 12.3$^a$ ($V$)&     &     &     &               \\
V948 Her     &  8       &    6-7/10     & $C$: TYC 2086-0950-1 &10.18$^a$ ($V$)& 3.5 & 3.6 & 3.3 & 1.7           \\
             &          &               & $K$: TYC 2086-1298-1 & 12.1$^a$ ($V$)&     &     &     &               \\
KR Mon       &  9       &  11/08-03/09  & $C$: TYC 4833-1488   & 12.6$^b$ ($B$)&     & 4.4 & 5.1 &               \\
             &          &        & $K$: USNO-A2.0 0825-05761727& 12.9$^c$ ($B$)&     &     &     &               \\
UZ Sge       &  11      &    7/09       & $C$: TYC 1626-1888-1 &10.93$^a$ ($V$)& 4.6 &     & 4.9 &               \\
             &          &               & $K$: TYC 1626-1275-1 & 11.12$^a$($V$)&     &     &     &               \\
\hline
\multicolumn{9}{l}{$^a$\citet{HO00}, $^b$\citet{EG92}, $^c$\citet{MO98}}
\end{tabular}}
\end{table}

\section{Light curve analysis}
\label{3}
%%%%general
The LCs of each system were analysed simultaneously using the \emph{PHOEBE} v.0.29d software \citep{PZ05} that follows the method of the 2003 version of the Wilson-Devinney (WD) code \citep{WD71,WI79,WI90}. All photometric points were given the same statistical weight (i.e. 1) and all of them were taken into account in the LC analysis. Due to the absence of spectroscopic mass ratios, the `$q$-search' method using a step of 0.1 was trialled in modes~2 (detached system), 4 (semi-detached system with the primary component filling its Roche Lobe) and 5 (semi-detached system with the secondary component filling its Roche Lobe) to find `photometric' estimates for the mass ratio $q_{\rm ph}$. Briefly, the $q$-search method can be described as follows: The $q$ value was kept fixed, while the other fitting parameters (see below for details) were adjusted. Therefore, the programme, for a given $q$ value, produced a sum of squared residuals $\sum res^{2}$. This procedure was repeated for a range of $q$ values and for different modes. The $q$ value corresponding to the minimum $\sum res^{2}$ (from all modes) was adopted as the most possible one. Then, this value was set as initial input and treated as free parameter in the subsequent analysis. The temperature values of the primaries were assigned values according to their spectral types by using the correlation given in the tables of \citet{CO00} and were kept fixed, while the temperatures of the secondaries $T_2$ were adjusted. The bolometric albedos, $A_1$ and $A_2$, and gravity darkening coefficients, $g_1$ and $g_2$, were set to the generally adopted values \citep{RU69,LU67,VZ24} for the given spectral types of the components. The linear limb darkening coefficients, $x_1$ and $x_2$, were taken from the tables of \citet{VH93}; the dimensionless potentials $\Omega_{1}$ and $\Omega_{2}$, the fractional luminosity of the primary component $L_{1}$ and the inclination $i$ of the system's orbit were set in the programme as adjustable parameters. Fig.~1 shows the $q$-search plot for each system for the finally adopted mode. Best-fit models, observed LCs and 3D plots are presented in Figs~2-3 with corresponding parameters given in Table~\ref{tab2}. The parameters' formal errors, given in parentheses, were directly calculated from the programme.

%%%%%%%%%%%%%%%%%%%%%%%%%%%%%%%%%%%%%%%%%%%%%%%%%%%%%%%%%%%%%%%%%%%%%%%%%%%%%%%%%%%%%%%%%%%%%%%%%%%%%%%%%%%%%%%%%%%%%%%%%%%Figure 1 - qs plots

\begin{figure}[h!]
\centering
\includegraphics[width=12cm]{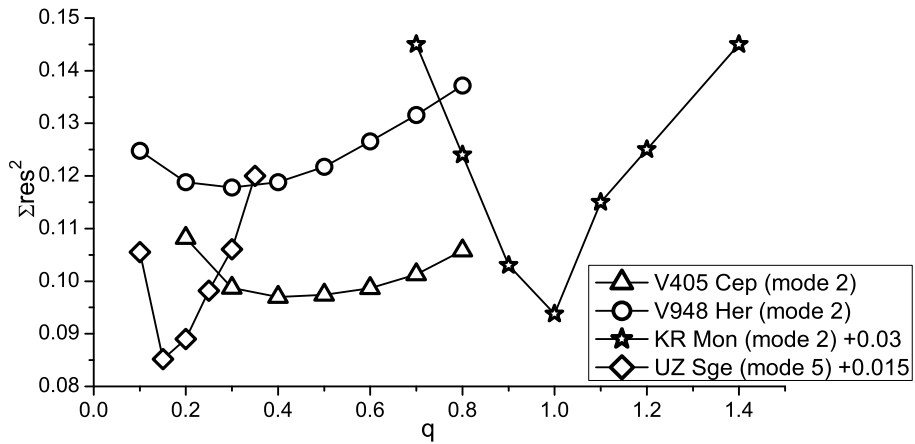}
\label{fig1}
\caption{Sum of squared residuals over a range of $q$ values for all systems.}
\end{figure}
%\vspace{5mm}

%%%%%%%%%%%%%%%%%%%%%%%%%%%%%%%%%%%%%%%%%%%%%%%%%%%%%%%%%%%%%%%%%%%%%%%%%%%%%%%%%%%%%%%%%%%%%%%%%%%%%%%%%%%%%%%%%%%%%%%%%%%Figures 2-3 - LC fittings & 3D Models

\begin{figure}[h!]
\centering
\begin{tabular}{c}
V405 Cep\\
\includegraphics[width=13.5cm]{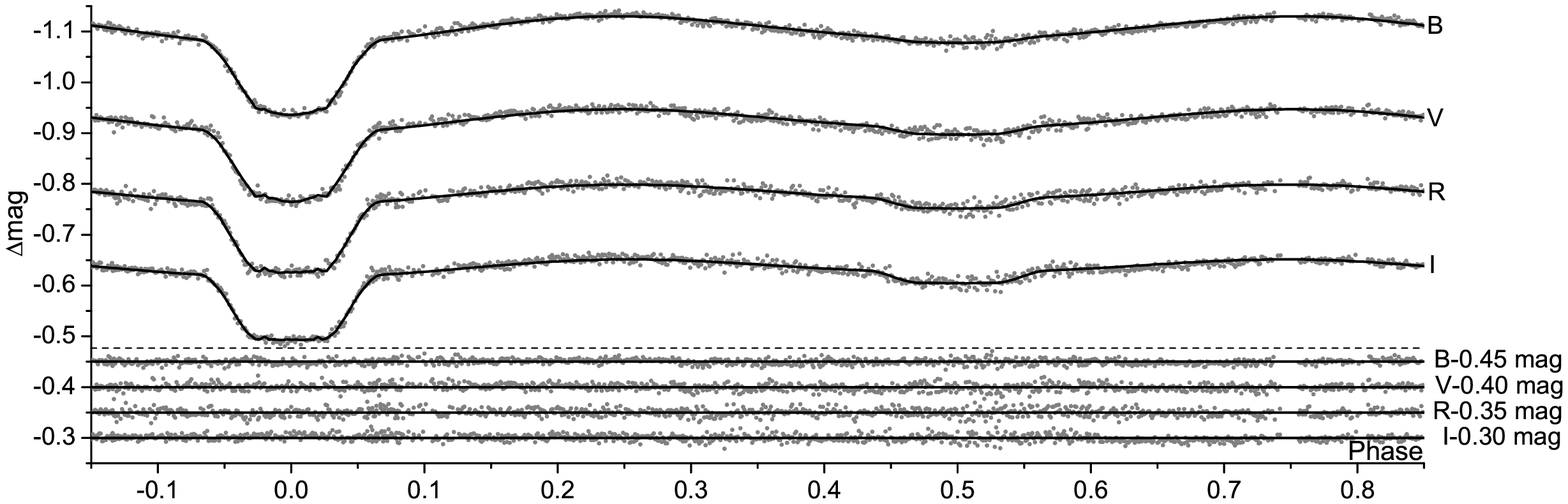}\\
V948 Her\\
\includegraphics[width=13.5cm]{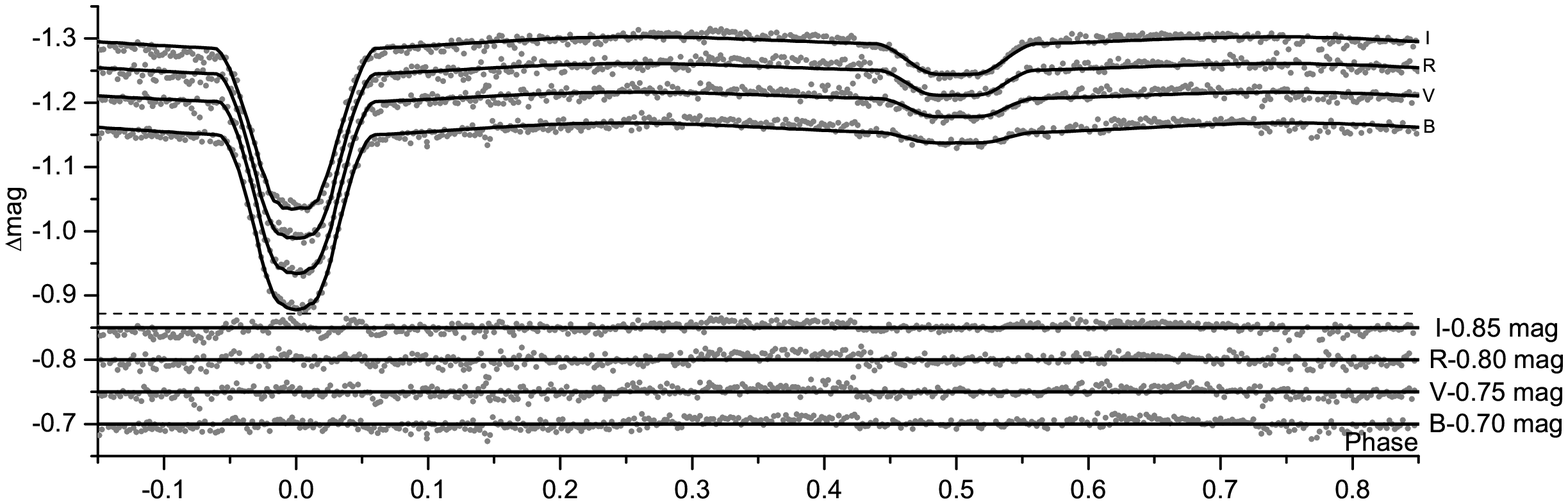}\\
KR Mon\\
\includegraphics[width=13.5cm]{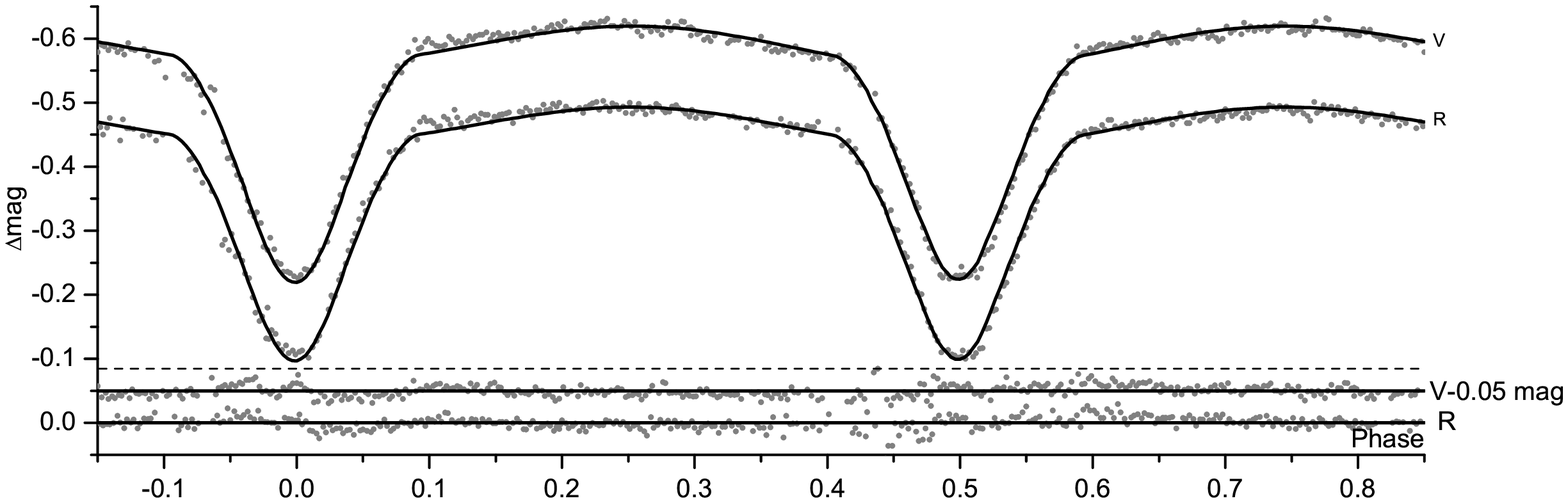} \\
UZ Sge\\
\includegraphics[width=13.5cm]{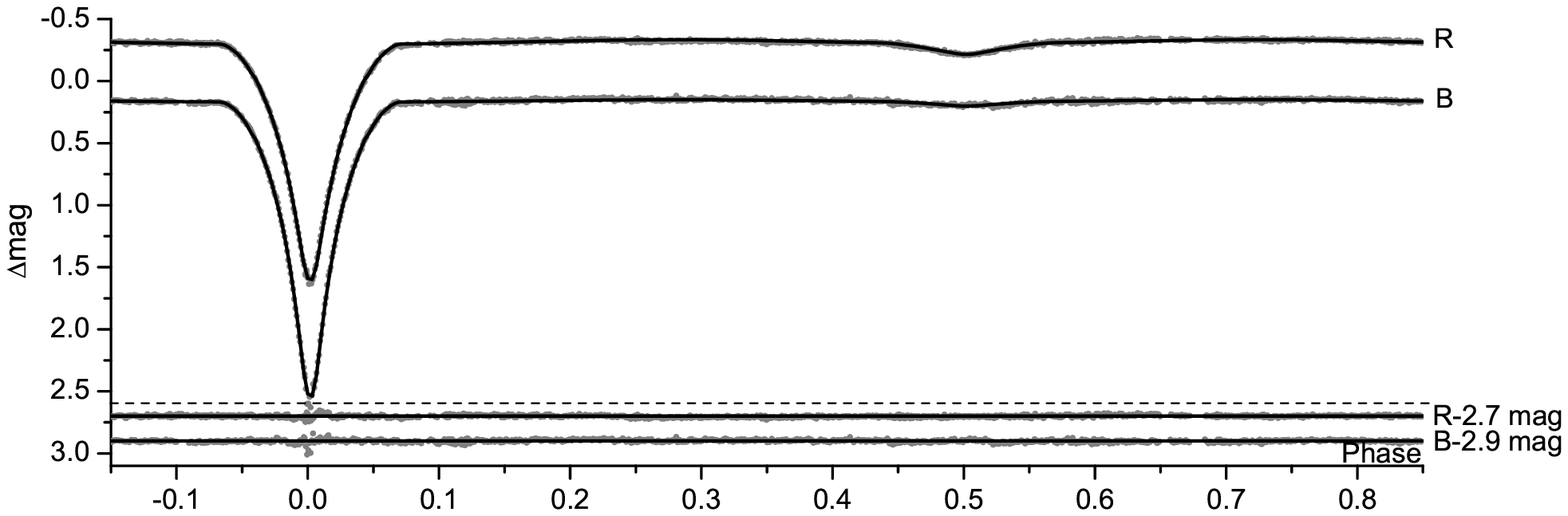} \\
\end{tabular}
\label{fig2}
\caption{Theoretical (solid lines) and observed (grey points) LCs and their respective residuals (below the dashed lines) for all systems.}
\end{figure}
%\vspace{5mm}

\begin{figure*}%[h!]
\centering
\begin{tabular}{ccc}
V405 Cep&V948 Her\\
\includegraphics[width=6cm]{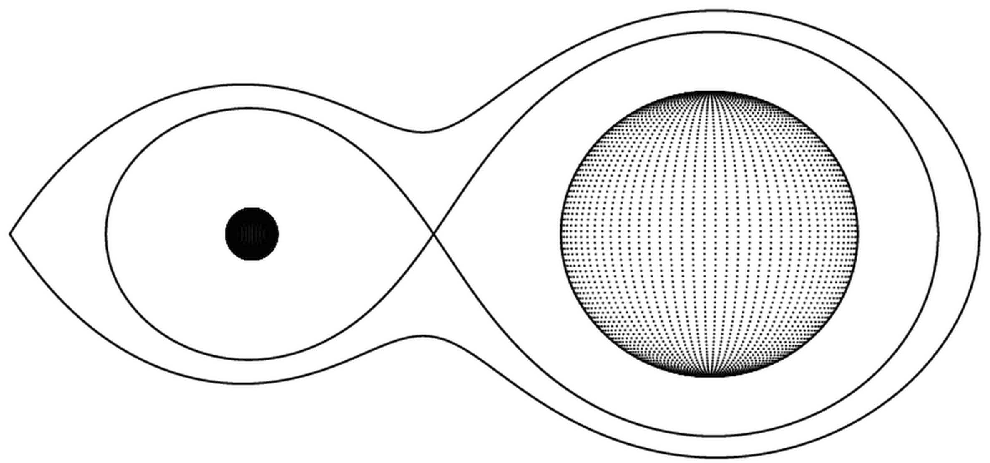}&\includegraphics[width=6cm]{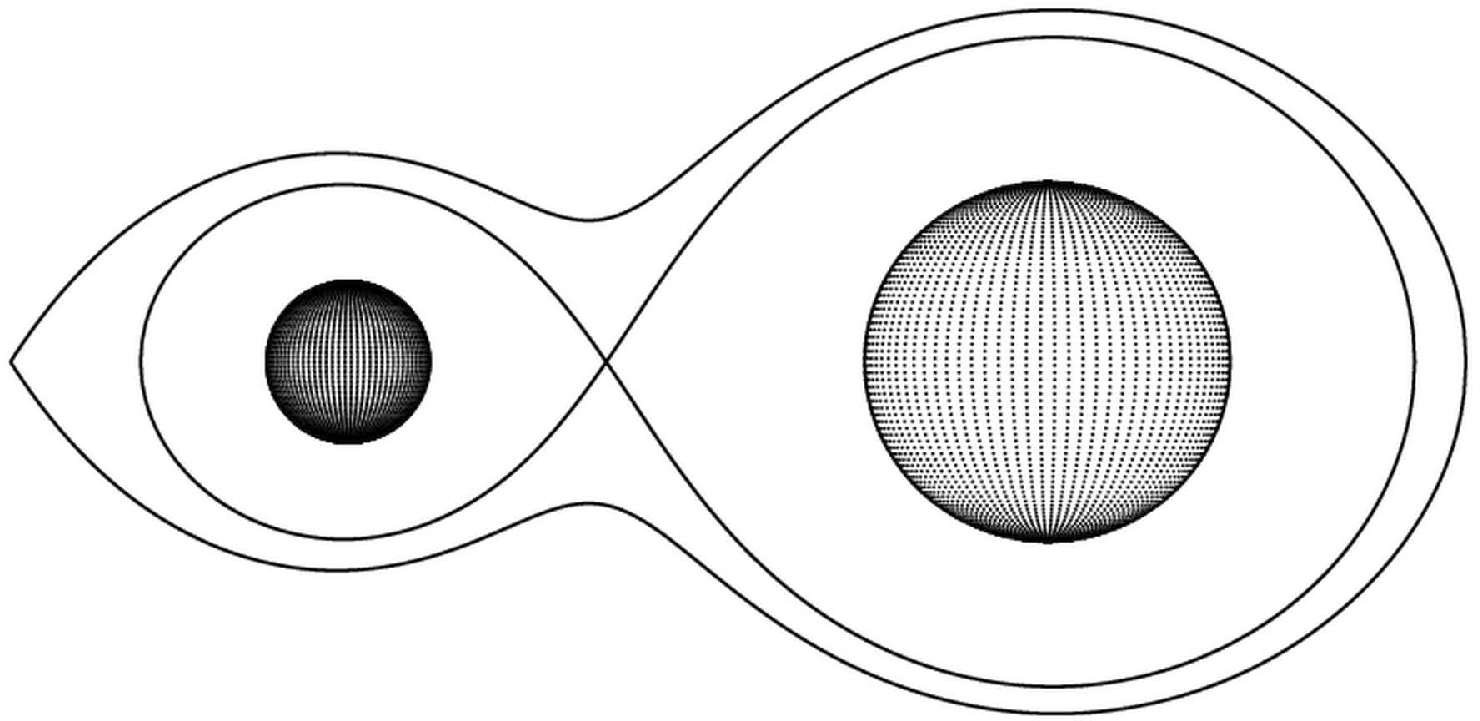}\\
KR Mon& UZ Sge\\
\includegraphics[width=6cm]{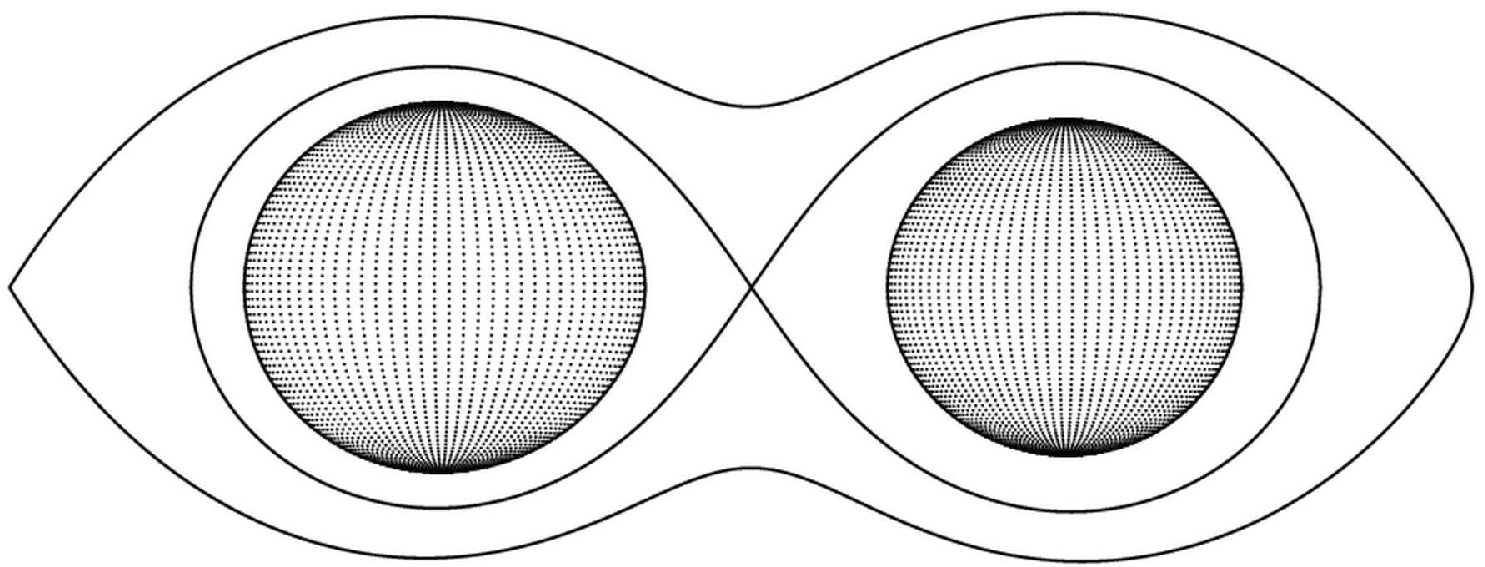}&\includegraphics[width=5cm]{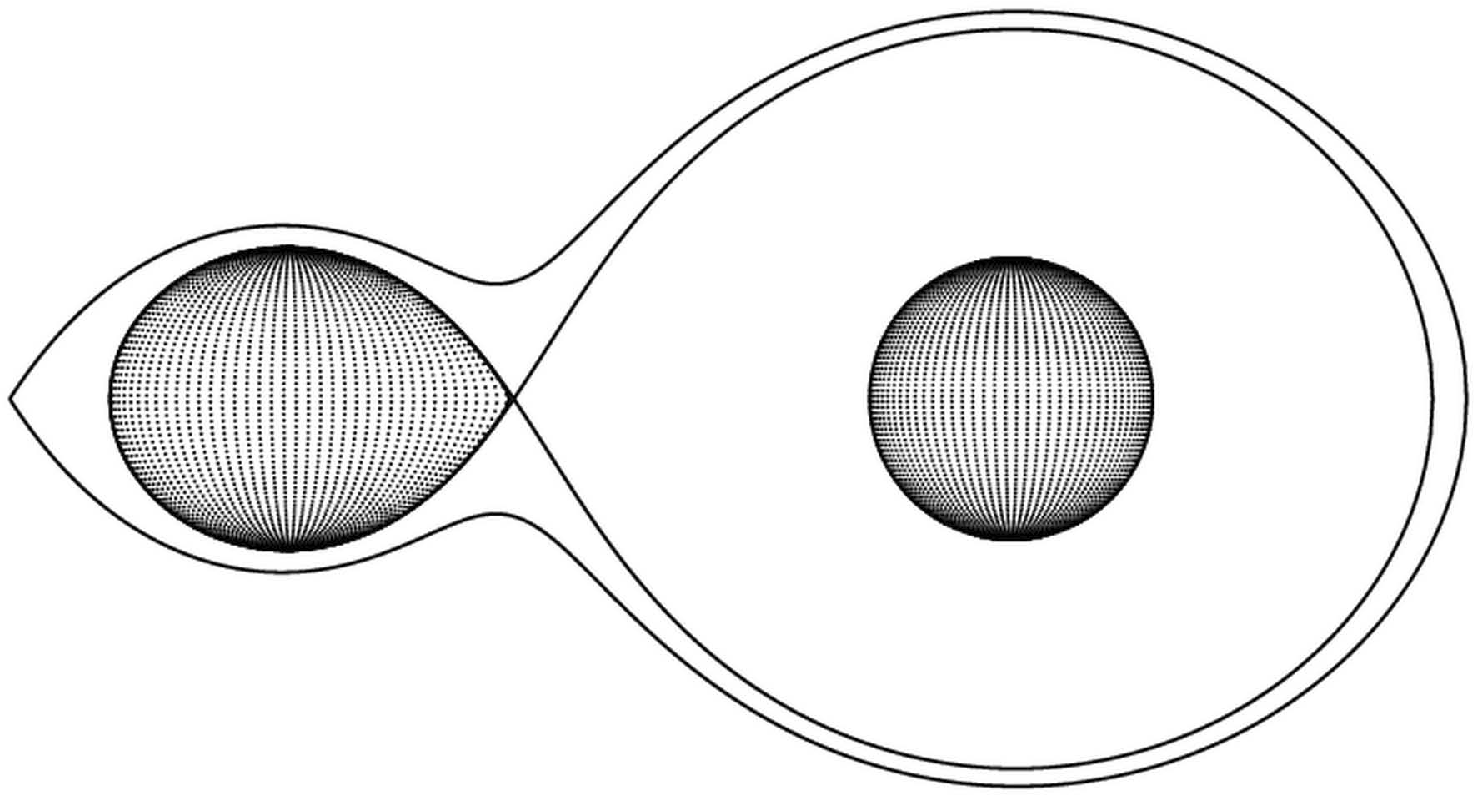}
\end{tabular}
\label{fig3}
\caption{3D plots at the phase 0.75 for all systems.}
\end{figure*}
%\vspace{5mm}

A value of 9000~K, typical for an A2 type star, was given to the primary component's temperature for V405~Cep. Mode 2 provided the best fit for the data, and the mass ratio, through the $q$-search method, turned out as $\sim$0.4.

The effective temperature of the primary component $T_1$ of V948~Her was set to 7000~K according to its spectral type (F2V). In the LC analysis, the minimum sum of weighted squared residuals ($\sum res^{2}$), was found in mode 2, where the $q$-search yielded a value of mass ratio around 0.3.

According to the literature (see Section~1), a G3V spectral type for the primary component of KR~Mon was suggested, therefore $T_1$ was given a value of 5700~K. The best fit was achieved in mode 2, where the mass ratio yielded a value of $\sim$1.

The most reliable spectral estimation was given by \citet{HA84}, where the spectral type of the system is referred as A3, therefore a value of 8700~K was assigned to the primary's temperature. Due to possible existence of a tertiary component, suggested by the cyclic orbital period changes \citep{LN08,ZA08}, the third light option $l_{3}$ was taken into account. The minimum $\sum res^{2}$ was found in mode 5, where the `$q$-search' method converged to a mass ratio value close to 0.2.

%%%%%%%%%%%%%%%%%%%%%%%%%%%%%%%%%%%%%%%%%%%%%%%%%%%%%%%%%%%%%%%%%%%%%%%%%%%%%%%%%%%%%%%%%%%%%%%%%%%%%%%%%%   Table 2 - The light curve solution parameters
\begin{table}[h!]
\centering
\caption{The light curve parameters.}
\label{tab2}
%\vspace{1mm}
\scalebox{0.95}{
\begin{tabular}{l cc cc cc cc}
\hline
Parameter	        &	\multicolumn{2}{c}{\textbf{V405 Cep}}&\multicolumn{2}{c}{\textbf{V948 Her}}&\multicolumn{2}{c}{\textbf{KR Mon}}&\multicolumn{2}{c}{\textbf{UZ Sge}}	\\
\hline	
                    &                                           \multicolumn{8}{c}{\textit{System related}}                                            \\
\hline								
Mode	            &  \multicolumn{2}{c}{Detached} &\multicolumn{2}{c}{Detached}	&\multicolumn{2}{c}{Detached}	  &\multicolumn{2}{c}{Semi-Detached}\\
$i$ [$^\circ$]      &	\multicolumn{2}{c}{83 (1)}	& \multicolumn{2}{c}{84.4 (6)}	&	\multicolumn{2}{c}{77.1 (1)}  &	\multicolumn{2}{c}{88.8 (1)}   \\
$q$ ($m_{2}/m_{1}$)	&  \multicolumn{2}{c}{0.36 (2)} & \multicolumn{2}{c}{0.27 (3)}	&	\multicolumn{2}{c}{0.97 (1)}  &	\multicolumn{2}{c}{0.14 (1)}   \\
\hline	
                    &                                           \multicolumn{8}{c}{\textit{Component related}}                                         \\
\hline	
\textit{Component:}	&      P	      &	      S	    &	      P	    &	      S	    &	        P	  &  	    S	  &	      P	      &	      S	       \\
\hline	
$T$ [K]	            &	  9000*	      &	4513 (90)	&	  7000*	    &	4310 (63)	&	    5700*	  &	   5705 (20)  &	   8700*      &	   4586 (60)   \\
$A$*	            &	   1	      &	      0.5	&	      1	    &	0.5	        &	   0.5	      &	    0.5	      & 	1	      &	     0.5	   \\
$g$*	            &	   1	      &	  0.32	    &	     1	    &	 0.32    	&	     0.32	  &	   0.32	      &	       1	  &	      0.32	   \\
$\Omega$            &	  3.6 (1)	  &	 4.8 (2)	&	 4.2 (1)	&	3.7 (2)	    &	  4.66 (1)	  &	  4.25 (1)	  &	   5.26 (3)	  &	      2.08	   \\
$x_{\rm B}$	        &	    0.540	  &	  0.935	    &	   0.595	&	 0.860	    &	    --	      &	     --	      &	     0.526	  &	      0.928	   \\
$x_{\rm V}$	        &	    0.456	  &	   0.789	&	   0.494	&	0.790	    &	    0.611	  &	      0.610	  &	      --	  &	         --	   \\
$x_{\rm R}$	        &	    0.379	  &	  0.676	    &	    0.417	&	    0.682	&	    0.526	  &	      0.525	  &	      0.395	  &	      0.687	   \\
$x_{\rm I}$	        &	    0.298	  &	  0.569	    &	    0.341	&	    0.567	&	        --	  &	     --	      &	     --	      &	     --	       \\
$(L/L_{\rm T})_{\rm B}$&  0.998 (1)	  &	0.002 (1)	&	0.991 (2)	&	0.009 (1)	&	      --	  &	     --	      &	  0.887 (2)	  &	    0.043 (1)  \\
$(L/L_{\rm T})_{\rm V}$&  0.994 (1)	  &	0.006 (1)	&	0.981 (2)	&	0.019 (1)	&	  0.446 (2)	  &	   0.554 (2)  &	      --	  &  	   --	   \\
$(L/L_{\rm T})_{\rm R}$&  0.991 (1)	  &	0.009 (1)	&	0.972 (2)	&	0.028 (1)	&	  0.446 (2)	  &	   0.554 (2)  &	   0.819 (2)  &	     0.098 (1) \\
$(L/L_{\rm T})_{\rm I}$&  0.988 (1)	  &	0.012 (1)	&	0.966 (2)	&	0.034 (1)	&	      --	  &	     --	      &	     --	      &	     --	       \\
\hline	
                    &                                           \multicolumn{8}{c}{\textit{Third light}}                                               \\
\hline	
$(L_3/L_{\rm T})_{\rm B}$&	\multicolumn{2}{c}{--}  &	\multicolumn{2}{c}{--}	    &	  \multicolumn{2}{c}{--}	  &	\multicolumn{2}{c}{0.070 (1)}  \\
$(L_3/L_{\rm T})_{\rm V}$&	\multicolumn{2}{c}{--}  &	\multicolumn{2}{c}{--}	    &	  \multicolumn{2}{c}{--}	  &	\multicolumn{2}{c}{--}	       \\
$(L_3/L_{\rm T})_{\rm R}$&	\multicolumn{2}{c}{--}  &	\multicolumn{2}{c}{--}	    &	  \multicolumn{2}{c}{--}	  &	\multicolumn{2}{c}{0.084 (1)}  \\
$(L_3/L_{\rm T})_{\rm I}$&	\multicolumn{2}{c}{--}  &	\multicolumn{2}{c}{--}	    &	  \multicolumn{2}{c}{--}	  &	\multicolumn{2}{c}{--}	       \\
\hline
$\sum res^{2}$      &   \multicolumn{2}{c}{0.092}   &  \multicolumn{2}{c}{0.106}	&	 \multicolumn{2}{c}{0.060}    & \multicolumn{2}{c}{0.067}      \\
\hline
\multicolumn{9}{l}{\textit{*assumed}, $L_{\rm T}=L_1+L_2+L_3$, $L_3=4\pi l_3$, \textit{P=Primary}, \textit{S=Secondary}}
\end{tabular}}
\end{table}

\section{Absolute parameters and evolutionary status}
\label{4}

\begin{figure}[h]
\centering
\begin{tabular}{cc}
\includegraphics[width=8cm]{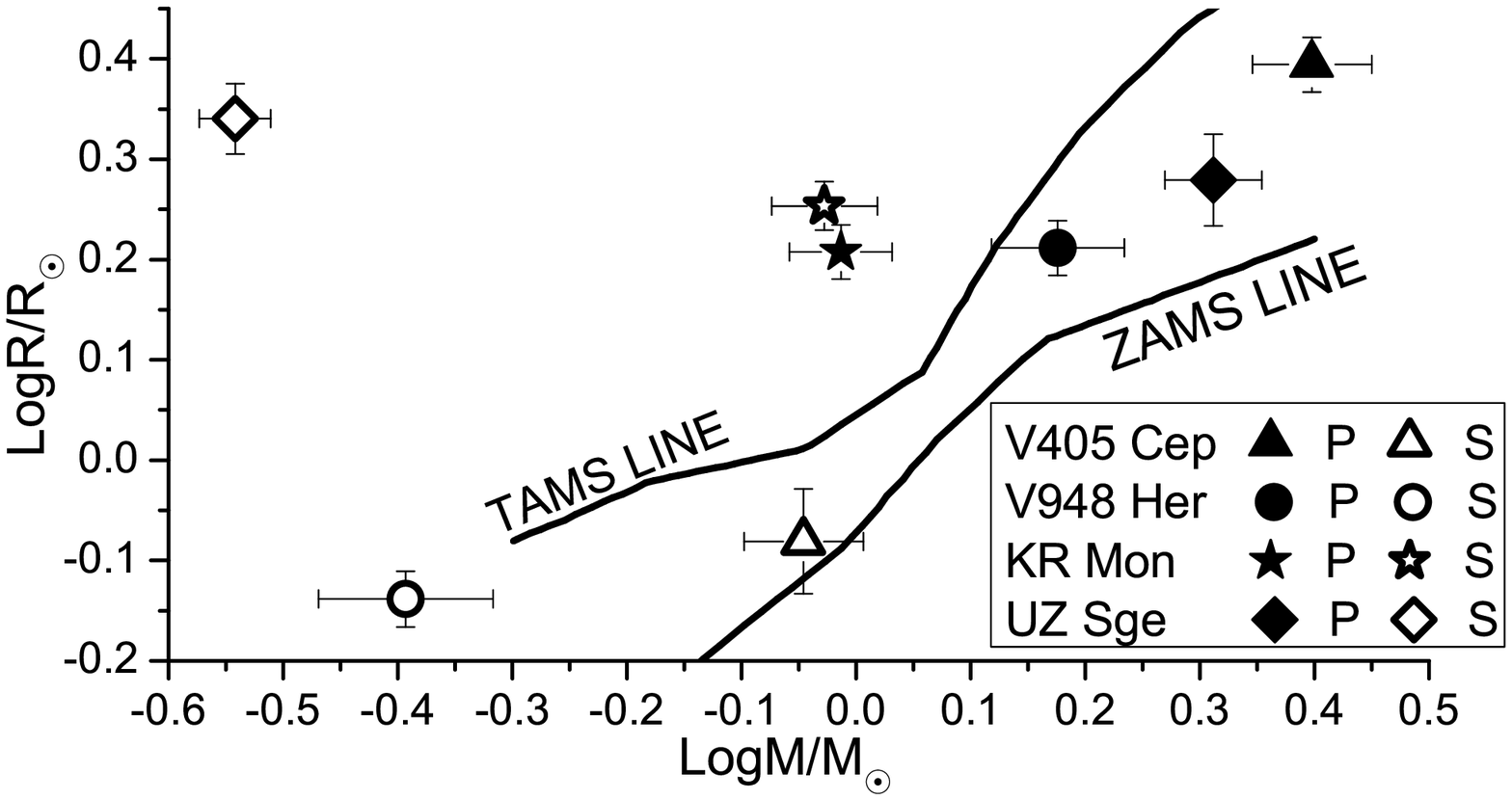}&\includegraphics[width=8cm]{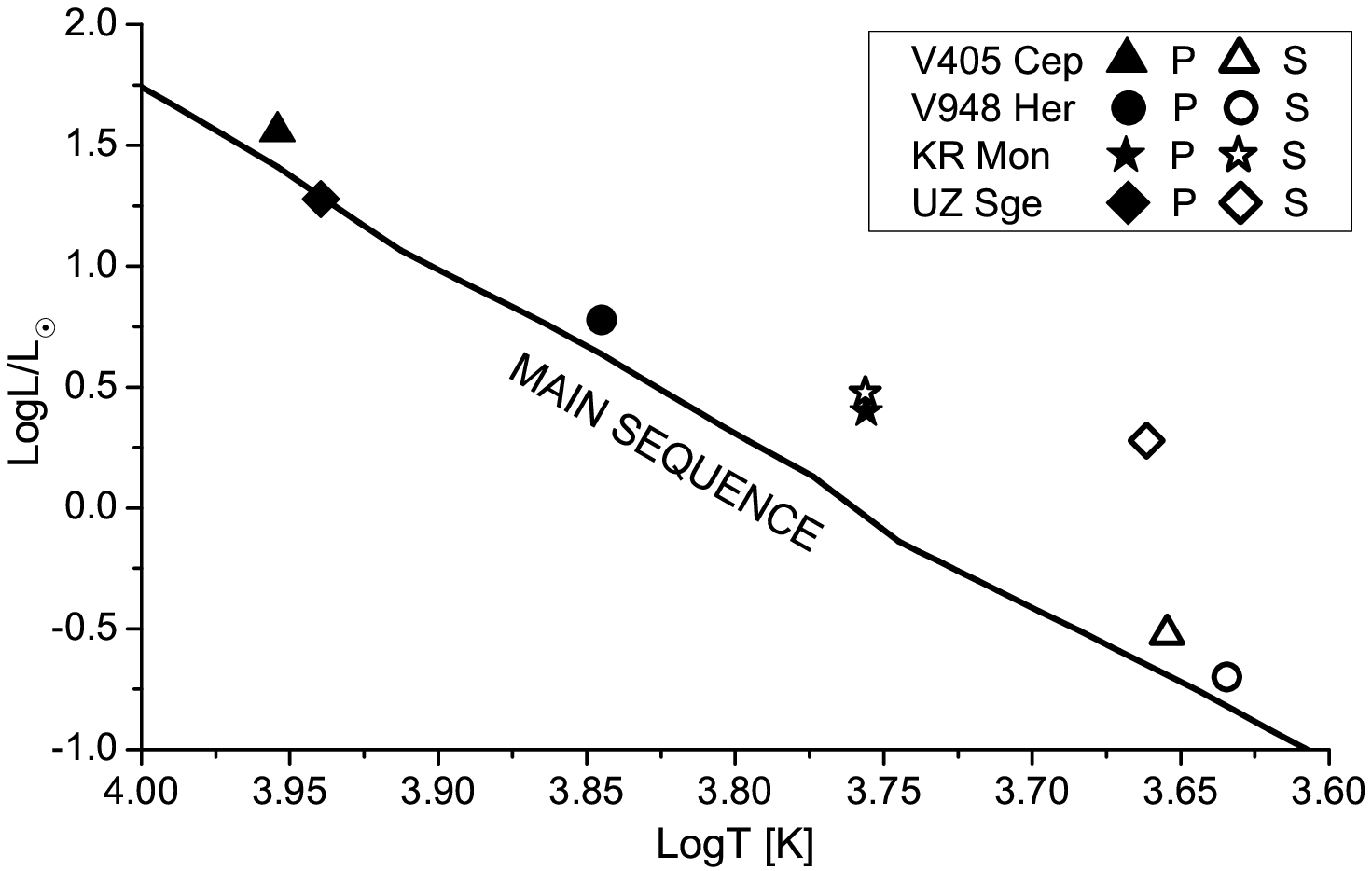}
\end{tabular}
\label{fig4}
\caption{The location of the components (open and filled symbols) of the systems in the $M-R$ (left) and $CM$ (right) diagrams. The indicators $P$ and $S$ refer to the primary and secondary components, respectively. The errors of the points in the $CM$ diagram are not shown due to scale.}
\end{figure}

Although no radial velocity curves exist for the systems, we can make fair estimates of their absolute parameters. These are listed in Table~\ref{tab3}. The mass of the primary component $M_{\rm P}$ of each system was assumed from its spectral type according to the tables of \citet{CO00} for Main Sequence (MS) stars, while masses of the secondaries followed from the adopted mass ratios given above (see Table \ref{tab2}). The semi-major axes $a$, used to calculate mean radii, then follow from Kepler's third law. A fair error of $M_{\rm P}$ was assumed to be $\sim10$\% of the mass value. The errors of the rest parameters were calculated based on the error propagation method and they are indicated in parentheses alongside with the adopted values. In Fig.~\ref{fig4}, the positions of the components of each system in the Mass-Radius ($M-R$) and Colour-Magnitude ($CM$) diagrams are illustrated.

The primary component of V405~Cep was found to be almost in the middle of Terminal Age Main Sequence (TAMS) and Zero Age Main Sequence (ZAMS) limits, while its secondary closer to ZAMS. Both components of V948~Her were found to be MS stars but closer to the TAMS edge. On the other hand, the components of KR~Mon are slightly evolved lying beyond the TAMS. The primary of UZ~Sge is a MS star and near to ZAMS, while its secondary is at the subgiant evolutionary stage. The TAMS and ZAMS lines were taken from \citet{NM03}, and the MS line for the $CM$ diagram from \citet{CO00}.

%%%%%%%%%%%%%%%%%%%%%%%%%%%%%%%%%%%%%%%%%%%%%%%%%%%%%%%%%%%%%%%%%%%%%%%%%%%%%%%%%%%%%%%%%%   Table 3 - The Absolute parameters of the components of the systems.
\begin{table}[h!]
\centering
\caption{Absolute parameters of the systems' components.}
\label{tab3}
%\scalebox{0.85}{
\begin{tabular}{l cc cc cc cc}
\hline
\textbf{Parameter}  &	\multicolumn{2}{c}{\textbf{V405 Cep}}&\multicolumn{2}{c}{\textbf{V948 Her}}&\multicolumn{2}{c}{\textbf{KR Mon}}&\multicolumn{2}{c}{\textbf{UZ Sge}}	\\
\hline
\textit{Component:}	&	  P	    &	    S	&	    P	&	S   	&	   P	&	  S	    &	    P	&	    S	\\
\hline
$M$ [M$_\odot$]	    &  2.5 (3)*	&	0.90 (4)&	1.5 (2)*&	0.40 (7)&0.97 (10)*	& 0.94 (10) &	2.1 (2)*&	0.29 (2)\\
$R$ [R$_\odot$]	    &	2.5 (1)	&	0.8 (1)	&	1.6 (1)	&	0.73 (3)&	1.6 (1)	&	1.8 (1)	&	1.9 (2)	&	2.2 (2)	\\
$\log g$ [cm/s$^2$]	&	4.0 (1)	&	4.6 (1)	&	4.2 (1)	&	4.3 (1)	&	4.0 (1)	&	3.9 (1)	&	4.2 (1)	&	3.2 (3)	\\
$a$ [R$_\odot$]	    &	2.1 (1)	&	5.9 (2)	&	1.3 (7)	&	4.9 (2)	&	2.9 (3)	&	3.0 (1)	&	1.2 (8)	&	8.6 (6)	\\
$L$ [L$_\odot$]	    &	38 (3)	&	0.3 (1)	&	6 (1)	&	0.2 (1)	&	2.5 (2)	&	3.0 (3)	&	19 (4)	&	1.9 (4)	\\
$M_{\rm bol}$ [mag] &  	0.8 (1)	&	6 (1)	&	2.9 (2)	&	7 (1)	&	3.8 (9)	&	4.0 (1)	&	1.6 (2)	&	4 (1)	\\
\hline
\multicolumn{9}{l}{\textit{*assumed, P=Primary, S=Secondary}}
\end{tabular}
\end{table}

%%%%%%%%%%%%%%%%%%%%%%%%%%%%%%%%%%%%%%%%%%%%%%%%%%%%%%%%%%%%%%%%%%%%%%%%%%%%%%%%%%%%%%%%%%%%%%%%%%%%%%%%%%%%%%%%%%%%%%%%%%%%%%%%% Figure 4 - M-R & CMD diagrams

\section{Search for pulsations}
\label{5}

Two of the systems of the present study, namely V405~Cep and V948~Her, were listed as candidates for containing a $\delta$ Sct component \citep{SO06}. In that work the candidacy criterion was the binary's spectral type (A-F). Since UZ~Sge has a spectral type in this range, we included it also in the frequency analysis. For this, the theoretical LCs of all filters were subtracted from the respective observed ones and a frequency search in the residuals ranging from 3 to 80~c/d \citep[typical for $\delta$~Sct stars]{BR00,SO06} was performed. The programme \emph{PERIOD04} v.1.2 \citep{LB05} that is based on classical Fourier analysis was used.

\begin{figure}[h]
\centering
\begin{tabular}{ccc}
                    V405 Cep                &                   V948 Her                 &                      UZ Sge              \\
\includegraphics[width=5cm]{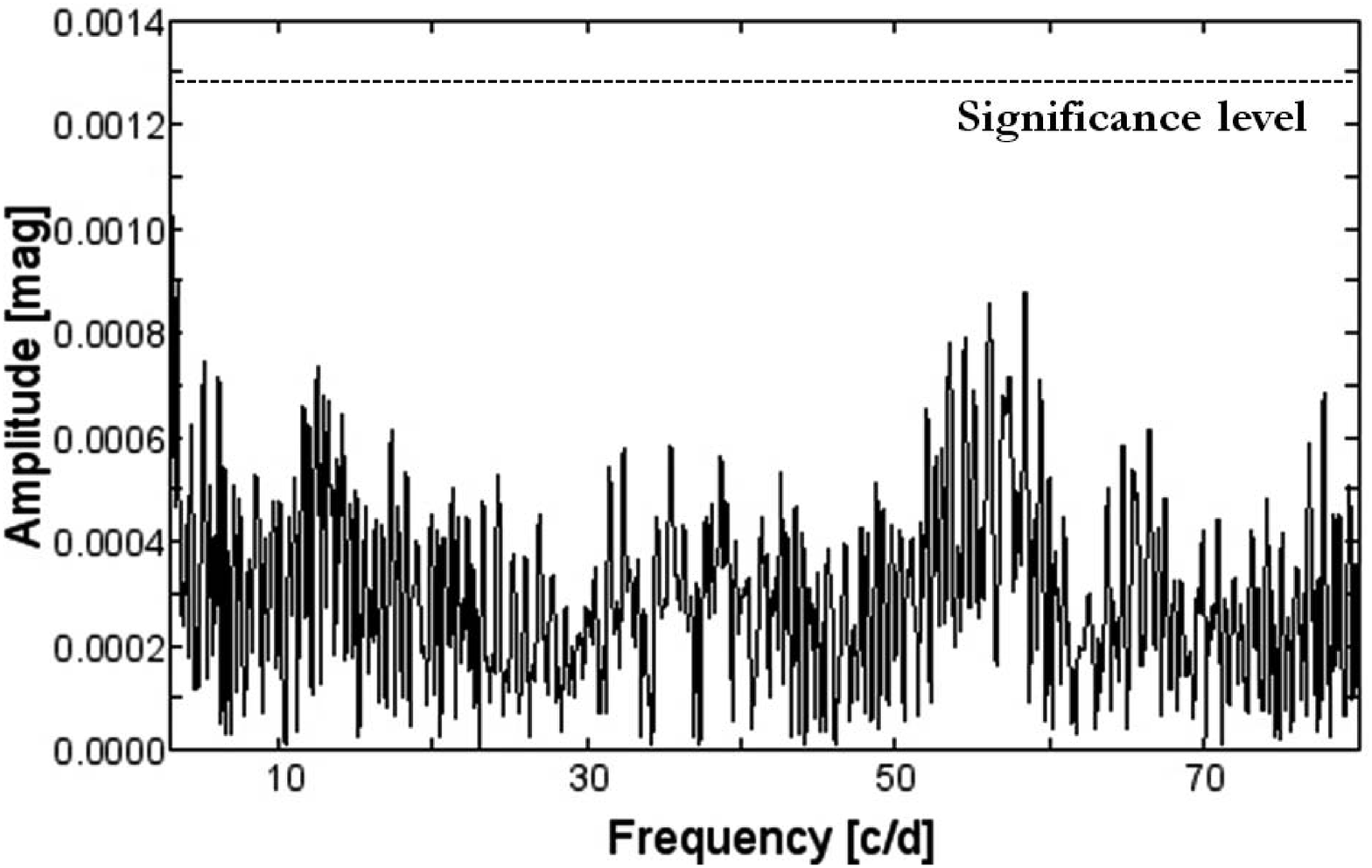}&\includegraphics[width=5cm]{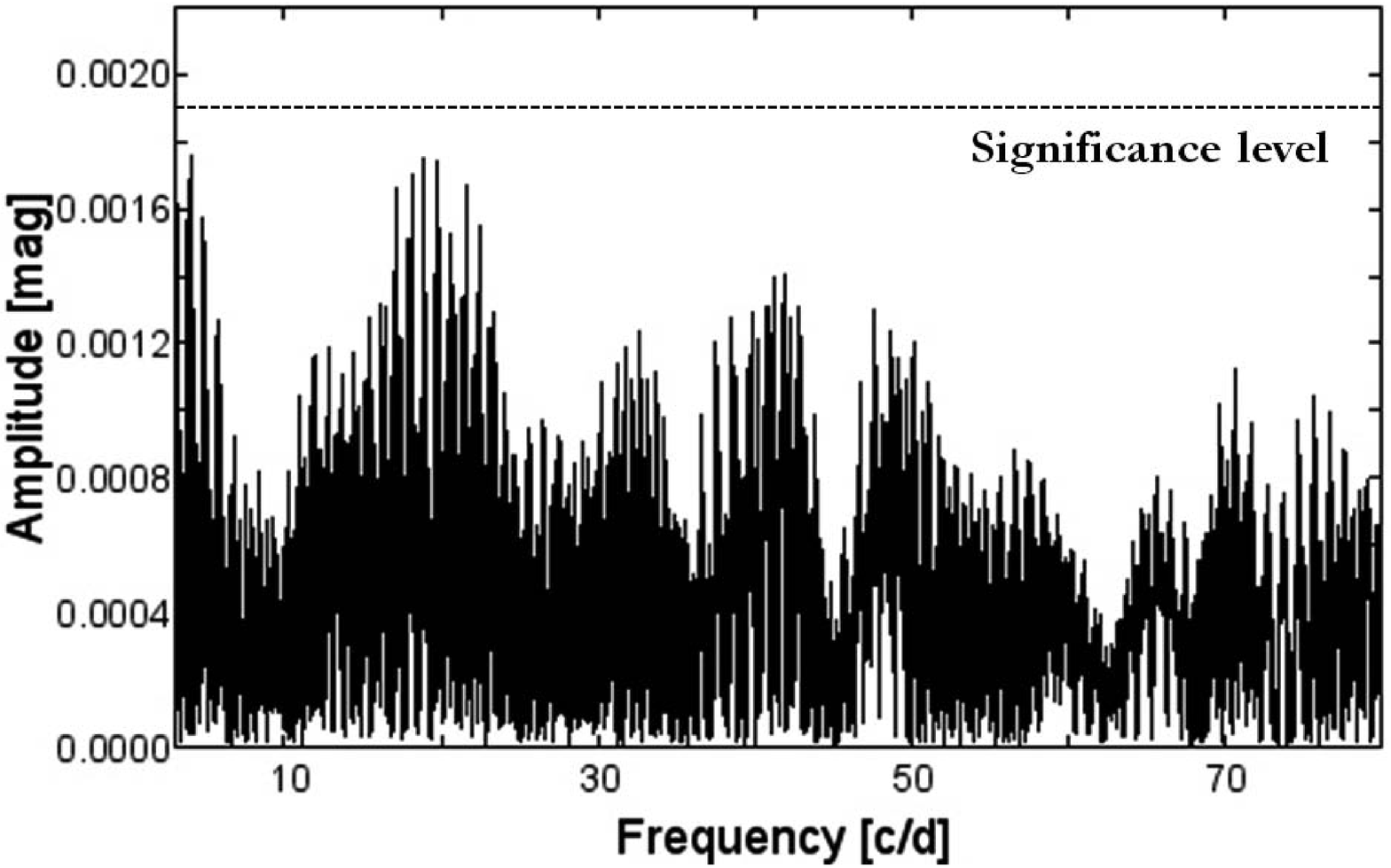}&\includegraphics[width=5cm]{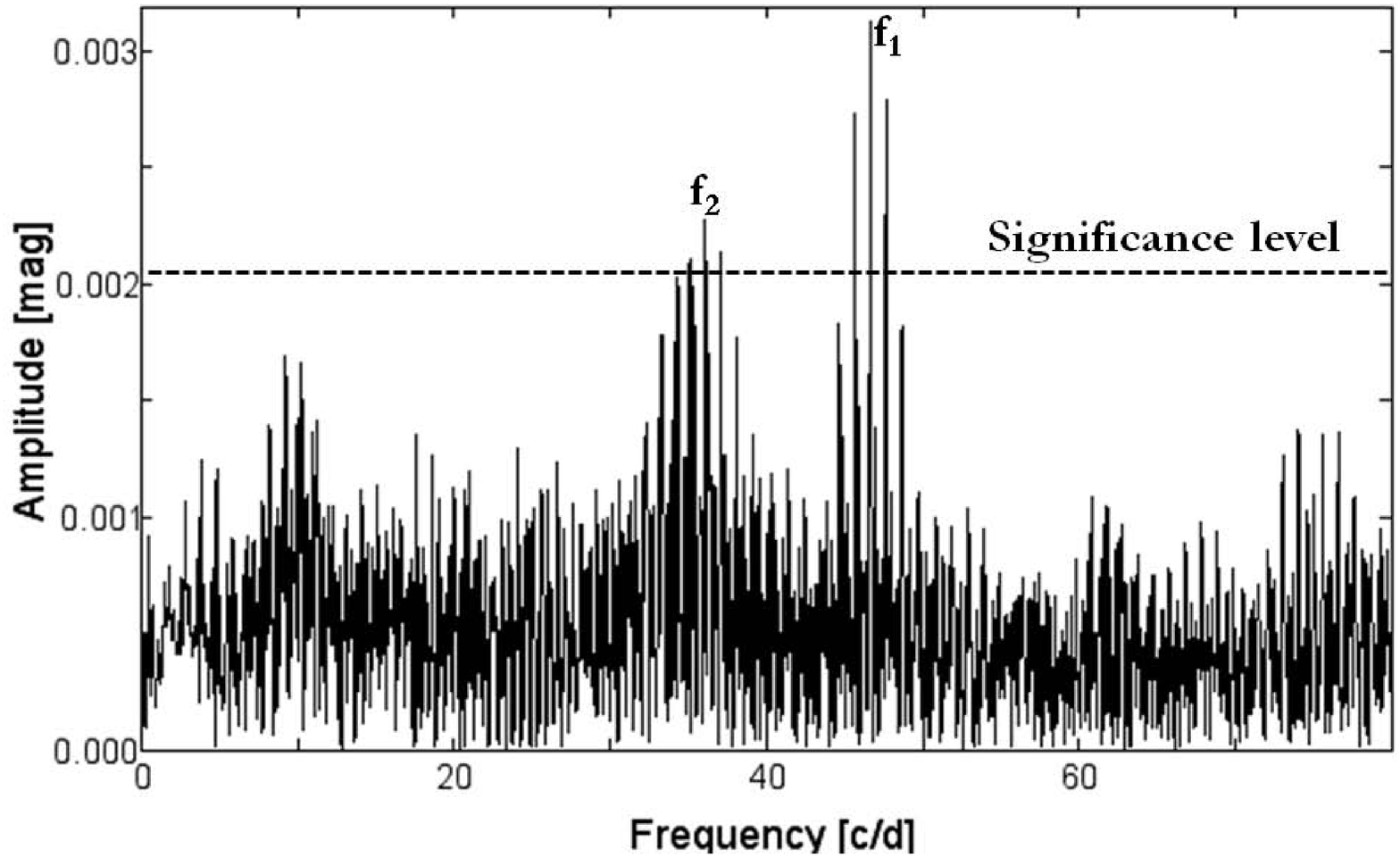}\\
\end{tabular}
\label{fig5}
\caption{Periodograms of all systems, where the significance level ($4\sigma$) is indicated. In the case of UZ~Sge the two frequencies overriding the critical detection limit are also indicated.}
\end{figure}

V405~Cep and V948~Her did not show convincing pulsational behaviour in the selected range of frequencies with a signal-to-noise ratio (S/N) higher than 4. On the contrary, for UZ~Sge we found two pulsation frequencies with the most dominant one at $\sim46.65$~c/d, which is well inside the frequency range of $\delta$ Sct type stars. The ratio of these frequencies ($f_2/f_1$) is $\sim0.75$, that is typical for the radial fundamental and first overtone modes. Table~\ref{tab4} contains the frequency analysis results for UZ~Sge, namely the frequency value $f$, the semi-amplitude $A$, the phase $\Phi$ and the $S/N$ value. In Fig.~\ref{fig5} are illustrated the amplitude spectra of all systems examined for pulsations.

%%%%%%%%%%%%%%%%%%%%%%%%%%%%%%%%%%%%%%%%%%%%%%%%%%%%%%%%%%%%%%%%%%%%%%%%%%%%%%%%%%%%%%%%%%%%%%%%%%%%%%%TABLE 4 ----Freqs for UZ~Sge
\begin{table}
\caption{Frequency analyses results for UZ~Sge.}
\label{tab4}
%\begin{minipage}{\columnwidth}
\centering
%\scalebox{0.98}{
\begin{tabular}{cc c ccc}
\hline										
filters  & 		 &	    $f$ 	&       $A$ 	&      $\Phi$   &  S/N  \\
         & 		 &	    [c/d]	&     [mmag]	&     $[^\circ$]&       \\
\hline											
																				
$B$     & $f_1$  &	 46.652 (6)	&	3.1 (3)	    &	354 (6)	    &  7.1	\\
        & $f_2$  &	 34.319 (9)	&	2.0 (3)	    &	195 (10)    &  4.0	\\
\hline											
$R$     &$f_1$   &	 46.650 (14)&	1.7 (3)	    &	18 (11)	    &  4.2	\\
        &$f_2$   &	 34.319 (9)	&	1.6 (3)	    &	197 (10)    &  4.0	\\
\hline																				
\end{tabular}
%\end{minipage}
\end{table}

%%%%%%%%%%%%%%%%%%%%%%%%%%%%%%%%%%%%%%%%%%%%%%%%%%%%%%%%%%%%%%%%%%%%%%%%%%%%%%%%%%%%%%%%%%%%%%%%%%%%%%%%%%%%%%%%%%%%%%%%%%%%%%%%% Figure 5 - periodograms

\section{Discussion and conclusions}
\label{6}

%%%%%%%%%%%%%%%%%%%%%%%%%%%%%%%%%%%%%%%%%%%%%%%%%%%%%%%%%%%%%%%%%%%%%%%%%%%%%%%%%%%%%%%%%Analyses brief
Complete LCs of the eclipsing systems V405~Cep, V948~Her, KR~Mon and UZ~Sge were obtained and analysed using well known modern techniques. The results were used to calculate the absolute parameters of their components and make an estimate of their present evolutionary status.

%V405 Cep
V405~Cep is a detached system with its primary component almost in the middle of ZAMS-TAMS limits, while its secondary was found to be closer to ZAMS. According to the mass values of these stars, it seems that the more massive component is more evolved that the less massive one. This situation is in agreement with the stellar evolutionary models. However, we notice that the stars show large radius difference, so, it is expected that the primary will fill its Roche lobe first as it evolves faster and, therefore, it will become a mass looser.

%V948 Her
For V948~Her it was found that its components are more evolved, reaching the TAMS, but they are well inside their Roche lobes. It is noticed that, although its components have a large difference in their masses, they are almost in the same evolutionary status. This fact comes in disagreement with our knowledge for stellar evolution, assuming a simultaneous birth of the components. This discrepancy could be explained with the scenario of past mass transfer from the present secondary to the primary, or less likely with a rapid mass loss (e.g. due to very strong stellar winds) of the secondary.

%KR Mon
On the other hand, the picture of KR~Mon is different. Its components have almost equal temperatures, masses and sizes, but none of them has reached its critical Roche volume so far. Both of them are slightly evolved and lie outside the TAMS limit. However, they are very close in filling their lobes, therefore they will probably come into contact in the future. Of course, the latter needs further justification with future observations, e.g. a period study showing if they are approaching each other.

%UZ Sge
UZ~Sge was found to be a classical Algol with the less massive and cooler component filling its Roche lobe. Its primary is close to ZAMS and was revealed as a $\delta$~Sct type pulsator, whose pulsating phase must have started relatively recently, if we consider both its dominant frequency value (i.e. fast pulsator) and its present evolutionary status \citep{LI12}. Therefore, according to the definition given by \citet{MK04} it can also be considered as oEA star (i.e. a classical Algol system including an oscillating MS component of A-F spectral type). The secondary component of the system is far beyond the MS edge and it must have been transferring a great part of its mass to the primary according to its present mass and evolutionary stage. Therefore, in the past it was very probably the more massive star of the system, evolved faster, reached its critical Roche limit and after that it started losing mass up to the present. However, according to the period analysis of the system \citep{LN08,ZA08} no secular period change that can be connected with the mass transfer process \citep{HI01} was found. Probably the system is at slow mass-accretion stage \citep{MK03} with a rate that cannot be detected with the current time coverage of minima timings. \citet{LN08} and \citet{ZA08} revealed also the possibility of a tertiary component's existence with a minimal mass of $\sim$0.7~M$_\odot$ and a mass function $f(m_3)=0.031$~M$_\odot$. From the present LC analysis we found a third light contribution of $L_{3,{\rm LC}}\sim$7.5\% to the total luminosity of the system. Assuming the MS nature of the third component, its luminosity based on the Mass-Luminosity relation for dwarf stars ($L\sim M^{3.5}$) can be derived. Given the absolute luminosities of the binary's members (see Table \ref{tab3}), we can estimate the expected light contribution ($L_{3,{\rm O-C}}$) of the potential third star by using the following formulae:
\begin{equation}
L_{3,{\rm O-C}} (\%)=100 \frac{M_{3,{\rm min}}^{3.5}}{L_1+L_2+M_{3,{\rm min}}^{3.5}}
\end{equation}

According to this hypothesis, we found that the third body's luminosity percentage is 1.4\%, that means less than the observed one. This difference, although it is large, it is based both on the minimal mass of the third body (coplanar orbit) and on its MS nature. Hence, if we use the mass function equation of the third body \citep[cf.][]{TO10}:

\begin{equation}
f(M_3) = \frac{1}{P_3^2}\left[\frac{173.145 A}{\sqrt{1 - e_3^2 \cos^2 \omega}} \right]^3 = \frac{(M_3 \sin i_3)^3}{(M_1+M_2+ M_3)^2}
\end{equation}
with the wide orbit's period $P_3$ in yr, and the LITE amplitude $A$ in days, we can see that, with $i_3\sim40^\circ$, we get $M_3\sim1.2$~M$_\odot$, a value satisfying the observed luminosity contribution. As an alternative, we can also assume that the third body is an evolved star providing more light than a MS star with the same mass. To sum up, we conclude that indeed there must be another component around the system, but its mass and nature remain open questions.

The mass ratio values for V405~Cep and V948~Her show deviation from the average of the `well known' detached binaries from the list of \citet{IB06}. However, that list contains only 74 cases of well studied detached binaries, therefore many deviations are expected to be found. Similar systems with relatively small mass ratio values are e.g. AL~Gem and UU~Leo \citep{LI11a}, EL~Vel \citep{ZA11}, GSC~4589-2999 \citep{LI11b} and AR~Lac \citep{IB06}. Moreover, the LCs of V405~Cep and V948~Her present total eclipses (both ones for V405~Cep and the secondary one for V948~Her), hence, the mass ratios can be considered of well determined using the `$q$-search' method \citep{TW05}. However, the error values of $q$ for these systems are not that realistic, given that their `$q$-search' diagrams show a shallow minimum (see Fig.~1), therefore, they can be considered only as formal ones. The parameters of KR~Mon and UZ~Sge are well inside the limits of detached and semi-detached systems, respectively \citep{IB06}.

%%%%%%%%%%%%%%%%%%%%%%%%%%%%%%%%%%%%%%%%%%%%%%%%%%%%%%%%%%%%%%%%%%%%%%%%%Suggestions+future work
Radial velocity measurements are certainly needed for deriving more accurate the absolute parameters of these close binaries and, therefore, obtaining a more realistic view of them.

\section*{Acknowledgments}

This work has been financially supported by the Special Account for Research Grants No 70/4/11112 of the National \& Kapodistrian University of Athens, Hellas.

\end{document}